\documentclass[runningheads]{svmult}

\usepackage{epsfig}
\usepackage{makeidx}   
\usepackage{graphicx}  
\usepackage{subeqnar}  
\usepackage{multicol}  
\usepackage{physprbb}  
\makeindex             

\newcommand{\BD}{\mbox{${\cal D}$}}
\newcommand{\lsim}{\mathrel{\rlap{\lower4pt\hbox{\hskip0pt$\sim$}}
\raise1pt\hbox{$<$}}}
\newcommand{\gsim}{\mathrel{\rlap{\lower4pt\hbox{\hskip0pt$\sim$}}
\raise1pt\hbox{$>$}}}
\newcommand{\sfrac}[2]{\mbox{\footnotesize $\frac{#1}{#2}$}}

\begin{document}
\title*{Diquarks and Density\vspace*{-2ex}}
\toctitle{Diquarks and Density}
%
\titlerunning{Diquarks and Density}
%
\author{M.B.~Hecht, C.D.~Roberts and S.M.~Schmidt}
\authorrunning{M.B.~Hecht, C.D.~Roberts and S.M.~Schmidt}
\institute{Physics Division, Argonne National Laboratory, Argonne, Illinois,
60439-4843, USA}

\maketitle              

\vspace*{-8ex}

\begin{abstract}
We describe aspects of the role that diquark correlations play in
understanding baryon structure and interactions.  The significance of
diquarks in that application motivates a study of the possibility that dense
hadronic matter may exhibit diquark condensation; i.e., quark-quark pairing
promoted by a quark chemical potential.  A Gorkov-Nambu-like gap equation is
introduced for QCD and analysed for $2$-colour QCD (QC$_2$D) and, in two
qualitatively different truncations, for QCD itself.  Among other interesting
features, we illustrate that QC$_2$D with massive fermions undergoes a
second-order transition to a superfluid phase when the chemical potential
exceeds $m_\pi/2$.  In the QCD application we illustrate that the
$\sigma:=-\langle \bar q q \rangle^{1/3} \neq 0$ phase, which determines the
properties of the mass spectrum at zero temperature and chemical potential,
is unstable with respect to the superfluid phase when the chemical potential
exceeds $\approx 2\,\sigma$, and that at this point the diquark gap is large,
$\approx \sigma/2$.  The superfluid phase survives to temperatures greater
than that expected in the core of compact stars.
\\[1ex]
To appear in the Proceedings of {\it Physics of Neutron Star Interiors}, a
workshop at ECT$\ast$, Trento, Italy, June-July/2000, Eds. D.~Blaschke,
N.K.~Glendenning and A.~Sedrakian.
\end{abstract}

\section{Diquarks}
A diquark is a bosonic quark-quark correlation, which is necessarily coloured
in all but $2$-colour QCD (QC$_2$D).  Therefore, in the presence of
colour-confinement, diquarks cannot be directly observed in a $N_c\geq 3$
colour gauge theory's spectrum.  Nevertheless evidence is accumulating that
suggests confined diquark correlations play an important role in hadronic
spectroscopy and interactions.

The first discussion of diquark correlations in literature addressing the
strong interaction is almost coincident with that of quarks
themselves~\cite{firstdq,lichtenberg}.  It was quickly realised that both
Lorentz scalar and vector diquarks, at least, are important for baryon
spectroscopy~\cite{L2} and, from a consideration of baryon magnetic
moments~\cite{L3}, that the diquark correlations are not pointlike.  This
latter point is still often overlooked, although with decreasing frequency
and now certainly without the imputation that it is a realistic
simplification.

The motivation for considering diquarks in the constituent-quark model is
that treating baryons directly as a three-body problem poses significant
challenges in anything other than a mean-field approach.  The task is much
simplified if two of the constituents can be replaced by a single degree of
freedom.  However, an obvious question is whether there is any sense in which
that replacement is more than just an expedient; i.e., a sense in which it
captures some important aspect of QCD's dynamics?  The answer is ``yes'' and
we now turn to explaining that.

A significant step toward a description of baryons in quantum field theory
can be identified in the realisation~\cite{regdq} that a large class of field
theoretical models of the strong interaction admit the construction of a
meson-diquark auxiliary-field effective action and thereby a description of
baryons as loosely-bound quark-diquark composites.  This is the class of
theories with a chiral symmetry preserving four-fermion interaction, which
includes, e.g., the Nambu--Jona-Lasinio model~\cite{njl} and the Global
Colour Model~\cite{gcm98}, that have been widely used in analysing low energy
strong interaction phenomena.

The picture of a baryon as loosely bound quark-diquark composite can also be
reached via a direct analysis of the bound state contributions to the three
quark scattering matrix.  The associated Schwinger function (Euclidean Green
function) is just that quantity whose large Euclidean-time behaviour yields a
baryon's mass in numerical simulations of lattice-QCD.  Considering the
colour structure of this Schwinger function, we focus on the Clebsch-Gordon
series for quarks in the fundamental representation of $SU_c(3)$:
\begin{equation}
3_c \otimes 3_c \otimes 3_c = (\bar 3_c   \oplus 6_c ) \otimes 3_c
= 1_c \oplus 8_c^\prime \oplus 8_c \oplus 10_c\,,
\end{equation}
from which it is clear that a {\em colour singlet} $3$-quark contribution is
only possible when two of the quarks are combined to transform according to
the {\em antitriplet}, $\bar 3_c$, representation.  This is the
representation under which antiquarks transform.

Single gluon exchange is repulsive in the $6_c$ channel but attractive in the
$\bar 3_c$ channel.  It is this feature that underpins the existence of the
meson-diquark bosonisation referred to above.  One way to see that is to
realise that the auxiliary field effective action obtained for any element of
the class of four-fermion interaction models provides a Lagrangian
realisation of the rainbow-ladder truncation of the Dyson-Schwinger equations
(DSEs)~\cite{cdragw}.  The rainbow-ladder truncation has been widely and
successfully employed in the study of meson spectroscopy and interactions,
see, e.g., Refs.~\cite{revbasti,revreinhard,pmqciv,cdrqciv}, and nonpointlike
colour-antitriplet diquark bound states exist in this truncation of the
quark-quark Bethe-Salpeter equation (BSE)~\cite{justindq}.  Hence they
provide a real degree of freedom to be used in the bosonisation.

At first sight the existence of colour-antitriplet diquark bound states in
these models, and in the rainbow-ladder truncation, appears to be a problem
because such states are not observed in the QCD spectrum.  However, as
demonstrated in Refs.~\cite{truncscheme,reinharddq}, this apparent lack of
confinement is primarily an artefact of the rainbow-ladder truncation.
Higher order terms in the quark-quark scattering kernel, the crossed-box and
vertex corrections, whose analogue in the quark-antiquark channel do not much
affect many of the colour singlet meson channels, act to ensure that the
quark-quark scattering matrix does not exhibit the singularities that
correspond to asymptotic (unconfined) diquark bound states.

Nevertheless, such studies with improved kernels, which do not produce
diquark bound states, do support a physical interpretation of the
``spurious'' rain\-bow-ladder diquark masses.  Denoting the mass in a given
diquark channel (scalar, pseudovector, etc.) by $m_{qq}$, then $\ell_{qq}:=
1/m_{qq}$ represents the range over which a true diquark correlation in this
channel can persist {\it inside} a baryon.  In this sense they are
``pseudo-particle'' masses that can be used to estimate which $\bar 3_c$
diquark correlations should dominate the bound state contribution to the
three quark scattering matrix, and hence which should be retained in deriving
and solving a Poincar\'e covariant homogeneous Fadde$^\prime$ev equation for
baryons.

The simple Goldstone-theorem-preserving rainbow-ladder kernel of
Ref.~\cite{conradsep} can be used to illustrate this point.  The model yields
the following calculated diquark masses (isospin symmetry is assumed):
\begin{equation}
\label{diquarkmasses}
\begin{array}{l|cccccccc}
(qq)_{J^P}           & (ud)_{0^+} & (us)_{0^+}  & (uu)_{1^+}& (us)_{1^+} 
        & (ss)_{1^+} & (uu)_{1^-} & (us)_{1^-} & (ss)_{1^-}\\\hline
 m_{qq}\,({\rm GeV}) & 0.74       & 0.88        & 0.95      & 1.05 
        & 1.13       & 1.47 & 1.53 & 1.64
\end{array}
\end{equation}
and the results are relevant because the mass ordering is characteristic and
model-independent, and lattice estimates, where
available~\cite{latticediquark}, agree with the masses tabulated here.
Equation~(\ref{diquarkmasses}) suggests that an accurate study of the nucleon
should retain the scalar and pseudovector correlations: $(ud)_{0^+}$,
$(uu)_{1^+}$, $(ud)_{1^+}$, $(dd)_{1^+}$, because for these diquarks $m_{qq}
\lsim m_N$, where $m_N$ is the nucleon mass, but may neglect other
correlations.  Furthermore, it is obvious from the angular momentum
Clebsch-Gordon series: $\sfrac{1}{2} \otimes\, 0 = \sfrac{1}{2}$ and $
\sfrac{1}{2}\otimes\,1 = \sfrac{1}{2} \oplus \sfrac{3}{2}$, that decuplet
baryons are inaccessible without pseudovector diquark correlations.  It is
interesting to note that $m_{(ud)_{0^+}}/m_{(uu)_{1^+}} = 0.78$~cf.\ $0.76 =
m_N/m_\Delta$ and hence one might anticipate that the presence of diquark
correlations in baryons can provide a straightforward explanation of the
$N$-$\Delta$ mass-splitting and other like effects.  These ideas were first
enunciated in Refs.~\cite{regfe,conradfe} and Ref.~\cite{reinhard} provides a
convincing demonstration of their efficacy.

Explicit calculations; e.g., Ref.~\cite{cdrqciv}, show that retaining only a
scalar diquark correlation in the kernel of the nucleon's Fadde$^\prime$ev
equation provides insufficient binding to obtain the experimental nucleon
mass: the best calculated value is typically $\sim 40$\% too large.  However,
with the addition of a pseudovector diquark it is easy to simultaneously
obtain~\cite{cdrqciv,reinhardDelta} the experimental masses of the nucleon
and $\Delta$.  Such calculations plainly verify the intuition that follows
from simple mass-counting: the pseudovector diquarks are an important but
subdominant element of the nucleon's Fadde$^\prime$ev amplitude (cf.\ the
scalar diquark) whilst being the sole component of the $\Delta$.

The presence of diquark correlations in baryons also affects the predictions
for scattering observables, which may therefore provide a means for
experimentally verifying the ideas described above.  For example, their
presence provides a simple explanation of the neutron's nonzero electric form
factor~\cite{cdrqciv}: charge separation arising from a heavy $(ud)$ diquark
with electric charge $\sfrac{1}{3}$ holding on to a relatively light,
electric charge $(-\sfrac{1}{3})$ $d$-quark.  And also a prediction for the
ratio of the proton's valence-quark distributions: $d/u:= d_v(x\to
1)/u_v(x\to 1)$, which can be measured in deep inelastic
scattering~\cite{wallyexp}.  In this case, diquark correlations with
differing masses in the nucleon's Fadde$^\prime$ev amplitude are an immediate
indication of the breaking of $SU(6)$ symmetry, hence $d/u\neq 1/2$.
Furthermore, if it were true that $m_{(qq)_{J^P}}\gg\!  m_{(ud)_{0^+}}$, for
all $J^P \neq 0^+$, then $d/u=0$.  However, as we have seen, in reality the
$1^+$ diquark is an important subdominant piece of the nucleon's
Fadde$^\prime$ev amplitude so that a realistic picture of diquarks in the
nucleon implies $0 < d/u < \frac{1}{2}$, with the actual value being a
sensitive measure of the proton's pseudovector diquark fraction.

\section{Superfluidity in Quark Matter}
We have outlined above the role and nature of diquark correlations in
hadronic physics at zero temperature and density, and emphasised that
diquarks are an idea as old as that of quarks themselves.  Another phenomenon
suggested immediately by the meson-diquark auxiliary-field effective action
is that of diquark condensation; i.e., quark-quark Cooper pairing, which was
first explored in this context using a simple version of the
Nambu--Jona-Lasinio model~\cite{kahana}.  A chemical potential promotes
Cooper pairing in fermion systems and the possibility that such diquark
pairing is exhibited in quark matter is also an old idea, early explorations
of which employed~\cite{bl84} the rainbow-ladder truncation of the quark DSE
(QCD gap equation).  That interest in this possibility has been renewed is
evident in a number of contributions to this
volume~\cite{krishna,thomas,fridolin}.  A quark-quark Cooper pair is a
composite boson with both electric and colour charge, and hence superfluidity
in quark matter entails superconductivity and colour superconductivity.
However, the last feature makes it difficult to identify an order parameter
that can characterise a transition to the superfluid phase: the Cooper pair
is gauge dependent and an order parameter is ideally describable by a
gauge-invariant operator.  This particularly inhibits an analysis of the
phenomenon using lattice-QCD.

\subsection{Gap Equation}
Studies of the gap equation that suppress the possibility of diquark
condensation show that cold, sparse two-flavour QCD exhibits a nonzero
quark-antiquark condensate: $\langle \bar q q \rangle \neq 0$.  If it were
otherwise then the $\pi$-meson would be almost as massive as the
$\rho$-meson, which would yield a very different observable world.  The quark
condensate is undermined by increasing $\mu$ and $T$, and there is a large
domain in the physical (upper-right) quadrant of the $(\mu,T)$-plane for
which $\langle\bar q q\rangle=0$: for the purpose of exemplification, that
domain can crudely be characterised as the set (see, e.g.,
Refs.~\cite{bastiscm,gregp,arne2}):
\begin{equation}
\{(\mu,T): \mu^2/\mu_c^2+T^2/T_c^2 > 1\,,\;\mu,T>0\,; \mu_c\sim
0.3\,\mbox{--}\,0.4\,{\rm GeV}, T_c\sim 0.15\,{\rm GeV}\}\,.
\end{equation}

Increasing temperature also opposes Cooper pairing.  However, since
increasing $\mu$ promotes it, there may be a (large-$\mu$,low-$T$)-subdomain
in which quark matter exists in a superfluid phase.  That domain, if it
exists, is unlikely to be accessible at the Relativistic Heavy Ion Collider,
because it operates in the high temperature regime, but may be realised in
the core of compact astrophysical objects, which could undergo a transition
to superfluid quark matter as they cool.  Possible signals accompanying such
a transition are considered in Refs.~\cite{krishna,fridolin}.

It was observed in Ref.~\cite{jacquesdq} that a direct means of determining
whether a $SU_c(N$) gauge theory supports scalar diquark condensation is to
study the gap equation satisfied by
\begin{equation}
\label{sinv}
\BD(p,\mu) := 
{\cal S}(p,\mu)^{-1} =\left(
\begin{array}{cc}
D(p,\mu) & \Delta^i(p,\mu)\, \gamma_5  \lambda_{\wedge}^i \\
 -\Delta^i(p,-\mu)\, \gamma_5  \lambda_{\wedge}^i
        & C D(-p,\mu)^{\rm t} C^\dagger
\end{array}\right).
\end{equation}
Here $T=0$, for illustrative simplicity and because temperature can only act
to destabilise a condensate, and, with $\omega_{[\mu]}= p_4+i\mu$,
\begin{equation}
\label{DCA}
D(p,\mu) = i \vec{\gamma}\cdot \vec{p}\, A(\vec{p}^2,\omega_{[\mu]}^2) +
B(\vec{p}^2,\omega_{[\mu]}^2 ) + i \gamma_4 \,\omega_{[\mu]}
\,C(\vec{p}^2,\omega_{[\mu]}^2 )\,;
\end{equation}
i.e., the inverse of the dressed-quark propagator in the absence of diquark
pairing.  (NB.\ For $\mu=0$, $A$, $B$ and $C$ are real functions.)  It is one
of the fundamental features of DSE studies that the existence of a nonzero
quark condensate: $\langle \bar q q \rangle \neq 0$, is signalled in the
solution of the gap equation by $B(\vec{p}\,^2,\omega_{[\mu]}^2 )\not\equiv
0$~\cite{revbasti}.

In Eq.~(\ref{sinv}), $\{\lambda_{\wedge}^i$, $i=1\ldots n^\wedge_c$,
$n^\wedge_c= N_c (N_c-1)/2\}$ are the antisymmetric generators of $SU_c(N_c)$
and $C=\gamma_2 \gamma_4$ is the charge conjugation matrix,
\begin{equation}
\label{CCmtx}
C\gamma_\mu^{\rm t}C^\dagger = -\gamma_\mu\,;\;
[C,\gamma_5]=0\,, 
\end{equation}
where $X^{\rm t}$ denotes the matrix transpose of $X$.  The key new feature
here is that diquark condensation is characterised by
$\Delta^i(p,\mu)\not\equiv 0$, for at least one $i$.  That is clear if one
considers the quark piece of the QCD Lagrangian density: $L[\bar q,q]$.  It
is a scalar and hence $L[\bar q,q]^{\rm t}= L[\bar q,q]$.  Therefore $L[\bar
q,q] \propto L[\bar q,q] + L[\bar q,q]^{\rm t}$, and it is a simple exercise
to show that this sum, and hence the action, can be re-expressed in terms of
a $2\times 2$ diagonal matrix using the bispinor fields
\begin{equation}
\label{QQCD}
Q(x) :=  \left(\begin{array}{c}
                        q(x)\\
                        \underline q(x):= \tau^2_f\, C\, \bar q^{\rm t}
                         \end{array} \right)\,,\;\;
\bar Q(x)  := \left(\begin{array}{cc}
                        \bar q(x)\;\;
                        \bar {\underline q}(x):= q^{\rm t} \,C\,\tau^2_f
\end{array} \right),
\end{equation}
where $\{\tau_f^i: i=1,2,3\}$ are Pauli matrices that act on the isospin
index.\footnote{We only consider theories with two light-flavours.
Additional possibilities open if this restriction is
lifted~\protect\cite{krishna,thomas}.}  It is plain upon inspection that a
nonzero entry: $d(x)\,\gamma_5$, in row-2--column-1 of this action-matrix
would act as a source for $q^{\rm t} \tau^2_f C \gamma_5 q$; i.e., as a
scalar diquark source.

It is plain now that the explicit $2\times 2$ matrix structure of
$\BD(p,\mu)$ in Eq.~(\ref{sinv}) exhibits a quark bispinor index that is made
with reference to $Q(x)$, $\bar Q(x)$.  This approach; i.e., employing a
``matrix propagator'' with ``anomalous'' off-diagonal elements, simply
exploits the Gorkov-Nambu treatment of superconductivity in fermionic
systems, which is explained in textbooks, e.g., Ref.~\cite{mattuck}.  It
makes possible a well-ordered treatment and makes unnecessary a truncated
bosonisation, which in all but the simplest models is a procedure difficult
to improve systematically.

The bispinor gap equation can be written in the form
\begin{equation}
\label{bispinDSE}
\BD(p,\mu)  =  \BD_0(p,\mu) 
+ \left( 
\begin{array}{cc}
\Sigma_{11}(p,\mu) & \Sigma_{12}(p,\mu)\\
\gamma_4\,\Sigma_{12}(-p,\mu)\,\gamma_4 & C \Sigma_{11}(-p,\mu)^{\rm t}
C^\dagger  
\end{array} \right),
\end{equation}
where the second term on the right-hand-side is just the bispinor self
energy.  Here, in the absence of a scalar diquark source term,
\begin{equation}
\BD_0(p,\mu) = (i\gamma\cdot p + m )\tau_Q^0 - \mu\,\tau_Q^3\,,
\end{equation}
with $m$ the current-quark mass, and the additional Pauli matrices:
$\{\tau_Q^\alpha,\alpha = 0,1,2,3\}$, act on the bispinor indices.  As we
will see, the structure of $\Sigma_{ij}(p,\mu)$ specifies the theory and, in
practice, also the approximation or truncation of it.

\section{Two Colours}
\label{sec:Nc2}
Two colour QCD (QC$_2$D) provides an important and instructive example.  In
this case
$ \Delta^i \lambda_\wedge^i =\Delta \tau^2_c $
in Eq.~(\ref{sinv}), with $\frac{1}{2}\vec{\tau_c}$ the generators of
$SU_c(2)$, and it is useful to employ a modified bispinor
\begin{eqnarray}
Q_2(x) &:= &\left(\begin{array}{c}
                         q(x)\\
\underline q_2:=\tau^2_c\,\underline q(x)  \end{array} \right),\;
\bar Q_2(x) :=  \left(\begin{array}{cc}
                        \bar q(x)\;\;
\bar {\underline q}_2(x):=\bar {\underline q}(x)\, \tau_c^2
\end{array} \right).
\end{eqnarray}
Embedding the additional factor of $\tau_c^2$ in this way makes it possible
to write the Lagrangian's fermion--gauge-boson interaction term as
\begin{equation}
\label{IQC2D}
\bar Q_2(x) \,\frac{i}{2} g \gamma_\mu \tau_c^k\tau_Q^0
\,Q_2(x)\,A_\mu^k(x) 
\end{equation}
because $SU_c(2)$ is pseudoreal; i.e.,
$ \tau^2_c\left(-\vec{\tau}_c\right)^{\rm t}\tau^2_c = \vec{\tau}_c$,
and the fundamental and conjugate representations are equivalent; i.e.,
fermions and antifermions are practically indistinguishable.  (That the
interaction term takes this form is easily seen using $L[\bar q,q]^{\rm t}=
L[\bar q,q]$.)

Using the pseudoreality of $SU_c(2)$ it can be shown that, for $\mu=0$ and in
the chiral limit, $m=0$, the general solution of the bispinor gap equation
is~\cite{jacquesdq}
\begin{equation}
\label{sinvgen2}
\BD(p) = i\gamma\cdot p\, A(p^2) + {\cal V}(-\mbox{\boldmath$\pi$})\, {\cal
M}(p^2)\,,\;\;
{\cal V}(\mbox{\boldmath$\pi$}) = 
\exp\left\{i \gamma_5\, \sum_{\ell=1}^5\,T^\ell\, \pi^\ell\right\}
= {\cal V}(-\mbox{\boldmath$\pi$})^{-1} \,,
\end{equation}
where $\pi^{\ell=1,\ldots,5}$ are arbitrary constants and 
\begin{equation}
\{T^{1,2,3}= \tau_Q^3 \otimes \vec{\tau_f},\, T^4= \tau_Q^1\otimes
\tau_f^0,\, T^5= \tau_Q^2\otimes\tau_f^0 \}\,,\; \{T^i,T^j\}=2 \delta^{ij}\,,
\end{equation}
so that ${\cal D}^{-1}$ is
\begin{equation} 
\label{Sp}
{\cal S}(p) = \frac{-i\gamma\cdot p A(p^2) + {\cal V}(\mbox{\boldmath$\pi$})
{\cal M}(p^2)} {p^2 A^2(p^2) + {\cal M}^2(p^2)}\,
:= \,-i\gamma\cdot p \,\sigma_V(p^2) + {\cal
V}(\mbox{\boldmath$\pi$})\,\sigma_S(p^2)\, .
\end{equation}
[To illustrate this, note that inserting
$\mbox{\boldmath$\pi$}=(0,0,0,0,-\frac{1}{4}\pi)$ produces an inverse
bispinor propagator with the simple form in Eq.~(\ref{sinv}).]

That the gap equation is satisfied for any constants $\pi^\ell$ signals a
vacuum degeneracy -- it corresponds to a multidimensional ``Mexican hat''
structure of the theory's effective potential, as noted in a related context
in Ref.~\cite{cr85}.  Consequently, if the interaction supports a mass gap,
then that gap describes a five-parameter continuum of degenerate condensates:
\begin{equation}
\label{cndst}
\langle \bar Q_2 
{\cal V}(\mbox{\boldmath$\pi$}) Q_2\rangle \neq 0\,,
\end{equation}
and there are 5 associated Goldstone bosons: 3 pions, a diquark and an
anti-diquark.  (Diquarks are the ``baryons'' of QC$_2$D.)  In the
construction of Eq.~(\ref{sinvgen2}) one has a simple elucidation of a
necessary consequence of the Pauli-G\"ursey symmetry of QC$_2$D; i.e., the
practical equivalence of particles and antiparticles.

For $m\neq 0$, the gap equation requires~\cite{jacquesdq}
${\rm tr}_{FQ} \left[ T^i {\cal V} \right] = 0$,
so that in this case only $\langle \bar Q_2 Q_2 \rangle \neq 0$ and now the
spectrum contains five degenerate but massive pseudo-Goldstone bosons.  This
illustrates that a nonzero current-quark mass promotes a quark condensate and
opposes diquark condensation.

For $\mu\neq 0$ the general solution of the gap equation has the form 
\begin{equation}
\label{BDmu} 
\BD(p,\mu)= \left(
\begin{array}{cc}
D(p,\mu) & \gamma_5\,\Delta(p,\mu)  \\
- \gamma_5\Delta^\ast(p,\mu) & C D(-p,\mu) C^\dagger
\end{array}
\right)\,.
\end{equation}
In the {\it absence} of a diquark condensate; i.e., for $\Delta \equiv 0$,
\begin{equation}
[U_B(\alpha), \BD(p,\mu)]=0\,, \; 
U_B(\alpha):= {\rm e}^{i \alpha \tau_Q^3 \otimes \tau_f^0}\,;
\end{equation}
i.e., baryon number is conserved in QC$_2$D.  This makes plain that the
existence of a diquark condensate dynamically breaks this symmetry.

\begin{figure}[t]
\centering{\ \epsfig{figure=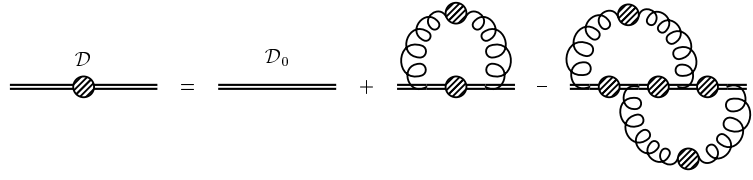,height=2.3cm}}\vspace*{0.5\baselineskip}

\caption{Illustration of the vertex-corrected gap equation, which is the
next-to-leading-order in the systematic, symmetry-preserving truncation
scheme of Ref.~\protect\cite{truncscheme}.  Retaining only the first two
diagrams on the right-hand-side yields the dressed-rainbow truncation.  Each
bispinor quark-gluon vertex is bare but the shaded circles mark quark and
gluon $2$-point functions that are dressed.  The corresponding truncation in
the relevant Bethe-Salpeter equations ensures the absence of diquark bound
states in the strong interaction spectrum.  (Adapted from
Ref.~\protect\cite{jacquesdq}.)
\label{dressed}}
\end{figure}

To proceed we choose to be explicit and employ the dressed-rainbow truncation
of the gap equation, see Fig.~\ref{dressed}, with a model for the Landau
gauge dressed-gluon propagator:
\begin{equation}
g^2 D_{\mu\nu}(k) = \left(\delta_{\mu\nu}-\frac{k_\mu k_\nu}{k^2}\right) 
                {\cal F}_2(k^2)\,,\;
{\cal F}_2(k^2)= \frac{64}{9}\pi^4\, {\hat \eta}^2\,\delta^4(k)\,.
\label{delta}
\end{equation}
This form for $g^2 D_{\mu\nu}(k)$ was introduced~\cite{mn83} for the
modelling of confinement in QCD but it is also appropriate here because the
string tension in QC$_2$D is nonzero, and that is represented implicitly in
Eq.~(\ref{delta}) via the mass-scale ${\hat \eta}$.

Using Eq.~(\ref{delta}) we obtain an algebraic dressed-rainbow gap equation
that, for $p^2=|\vec{p}|^2 + p_4^2 =0$, reads:
\begin{eqnarray}
A - 1 & = & \frac{1}{2}\,{\hat \eta}^2 \,K\,\left\{
                A \,(B^{\ast 2} - C^{\ast 2} \mu^2)
                + A^\ast \,|\Delta|^2 \right\}\,,\\
(C - 1)\,\mu & = & \frac{\mu}{2}\,{\hat \eta}^2 \,K\,\left\{ C\,(B^{\ast 2} -
                C^{\ast 2} \mu^2) -C^\ast\,|\Delta|^2 \right\}\,,\\
\label{Bmeqn}
B - m & = & {\hat \eta}^2 \,K\,\left\{
                B\,(B^{\ast 2} - C^{\ast 2} \mu^2)
                + B^\ast \,|\Delta|^2 \right\}\,,\\
\Delta & = & {\hat \eta}^2 \,K\, \left\{
                \Delta\,(|B|^2 + |C|^2 \mu^2)
                +\Delta \,|\Delta|^2 \right\}\,,
\end{eqnarray}
with 
$K^{-1}  =  |B^2-C^2\mu^2|^2 + 2 |\Delta|^2 (|B|^2+ |C|^2 \mu^2) +
|\Delta|^4\,. $
These equations possess a $B \leftrightarrow \Delta$ symmetry when
$(m,\mu)=0$, which is a straightforward illustration of the vacuum degeneracy
described above using the matrix ${\cal V}(\mbox{\boldmath$\pi$})$.  (Recall
that for $\mu=0$, $A$, $B$ and $C$ are real functions.)  They also exemplify
the general result that $\Delta$ is real for all $\mu$.  Another exemplary
result follows from a linearisation in $\mu^2$: $\mu\neq 0$ acts to promote a
nonzero value of $\Delta$ but oppose a nonzero value of $B$; i.e., a nonzero
chemical potential plainly acts to promote Cooper pairing at the expense of
$\langle \bar q q\rangle$.

For $(m,\mu)=0$ the solution of the dressed-rainbow gap equation obtained
using Eq.~(\ref{delta}) is:
\begin{eqnarray}
\label{ACeqn}
 & A(p^2) = C(p^2)  &=\left\{\begin{array}{ll} 2, & p^2 < \frac{{\hat
\eta}^2}{4}\\ \frac{1}{2}\left( 1 + \sqrt{1 + \frac{2{\hat
\eta}^2}{p^2}}\right), & {\rm otherwise}\;,
\end{array}\right.\\
{\cal M}^2(p^2)   & :=  B^2(p^2) + \Delta^2(p^2)    & =\left\{\begin{array}{ll}
{\hat \eta}^2 - 4 p^2, & p^2 < \frac{{\hat \eta}^2}{4}\\
0,              & {\rm otherwise}\;.
\end{array}\right.
\end{eqnarray}
As we have already mentioned, the dynamically generated mass function, ${\cal
M}(p^2)$, is tied to the existence of quark and/or diquark condensates, which
can be illustrated by noting that $(B=0, \Delta\neq 0)$ corresponds to
$\mbox{\boldmath$\pi$}=(0,0,0,0,\frac{1}{2}\pi)$ in Eq.~(\ref{cndst}); i.e.,
$ \langle \bar Q_2 i\gamma_5\tau_Q^2 Q_2\rangle \neq 0 $,
while $(B\neq 0, \Delta= 0)$ corresponds to
$\mbox{\boldmath$\pi$}=(0,0,0,0,0)$; i.e., 
$ \langle \bar Q_2 Q_2\rangle \neq 0 $.

The usual chiral, $SU_A(2)$, transformations are realised via
\begin{equation}
\BD(p,\mu) \to V(\vec{\pi})\, \BD(p,\mu) \,V(\vec{\pi}) \,,
\;\; V(\vec{\pi}):= {\rm e}^{i\gamma_5 \vec{\pi}\cdot\vec{T}}\,,\;
\vec{\pi}=(\pi^1,\pi^2,\pi^3)\,,
\end{equation}
and therefore, since the anticommutator $\{\vec{T},T^{4,5}\}=0$, a diquark
condensate does not dynamically break chiral symmetry.  On the other hand,
since $[\mbox{\boldmath $1$},\vec{T}]=0$, a quark condensate does dynamically
break chiral symmetry.  

In addition, and of particular importance, is the feature that in {\em
combination} with the momentum-dependent vector self energy the
momentum-dependence of ${\cal M}(p^2)$ ensures that the dressed-quark
propagator does not have a Lehmann representation and hence can be
interpreted as describing a confined
quark~\cite{cdragw,revbasti,revreinhard}.  The interplay between the scalar
and vector self energies is the key to this realisation of confinement.  The
qualitative features of this simple model's dressed-quark propagator have
been confirmed in recent lattice-QCD simulations~\cite{latticequark} and the
agreement between those simulations and more sophisticated DSE studies is
semi-quantitative \cite{pmqciv}.

In the steepest descent (or stationary phase) approximation the contribution
of dressed-quarks to the thermodynamic pressure is 
\begin{equation}
\label{pSigma}
p_{\Sigma}(\mu,T) = \frac{1}{2\beta V}\left\{ {\rm TrLn}\left[\beta
{\cal S}^{-1}\right] - \frac{1}{2}{\rm Tr}\left[\Sigma\,{\cal S}\right]\right\},
\end{equation}
where $\beta = 1/T$, and ``Tr'' and ``Ln'' are extensions of ``tr'' and
``$\ln$'' to matrix-valued functions.

The MIT Bag Model pictures the quarks in a baryon as occupying a spatial
volume from which the nontrivial quark-condensed vacuum (scalar-field) has
been expelled.  Therefore, as observed in Refs.~\cite{cr85}, the bag constant
can be identified as the pressure difference between the $\langle \bar q q
\rangle\neq 0$ vacuum, the so-called Nambu-Goldstone phase in which chiral
symmetry is dynamically broken, and the chirally symmetric no-condensate
alternative, which is called the Wigner-Weyl vacuum.  That difference is
given by
\begin{equation}
{\cal B}_B(\mu) := p_{\Sigma}(\mu,{\cal S}[B,\Delta=0]) -
p_{\Sigma}(\mu,{\cal S}[B=0,\Delta=0])\,,
\end{equation}
and it is, of course, $\mu$-dependent because the vacuum {\em evolves} with
changing $\mu$.  ${\cal B}_B$ also evolves with temperature and this
necessary $(\mu,T)$-dependence of the bag constant can have an important
effect on quark star properties; e.g., reducing the maximum supportable mass
of a quark matter star, as discussed in Ref.~\cite{quarkstar1}.

If we define, by analogy,
\begin{equation}
{\cal B}_\Delta(\mu) := p_{\Sigma}(\mu,{\cal S}[B=0,\Delta]) -
p_{\Sigma}(\mu,{\cal S}[B=0,\Delta=0])\,,
\end{equation}
then the relative stability of the quark- and diquark-condensed phases is
measured by the pressure difference
\begin{equation}
\label{deltaP}
\delta p(\mu) := {\cal B}_\Delta(\mu)  - {\cal B}_B(\mu)\,.
\end{equation}
For $\delta p(\mu) > 0$ the diquark condensed phase is favoured.

At $(m=0,\mu=0)$, $\delta p = 0$, with
\begin{equation}
{\cal B}_B(0) = {\cal B}_\Delta(0) = (0.092\,\hat\eta)^4\,.
\end{equation}
This equality is a manifestation of the vacuum degeneracy identified above in
connection with the matrix ${\cal V}(\mbox{\boldmath$\pi$})$.  However,
\begin{equation}
\mbox{with}\; m=0\,, \delta p > 0 \;\mbox{for all}\; \mu>0\,,
\end{equation}
which means that the Wigner-Weyl vacuum is unstable with respect to diquark
condensation for all $\mu> 0$~\cite{jacquesdq} and that the superfluid phase
is favoured over the Nambu-Goldstone phase.

Now, although the action for the $\mu\neq 0$ theory is invariant under
\begin{equation}
 Q_2 \to U_B(\alpha) \,Q_2\,,\; \bar Q_2 \to \bar Q_2\,U_B(-\alpha) \,, 
\end{equation}
which is associated with baryon number conservation, the diquark condensate
breaks this symmetry:
\begin{equation}
\langle \bar Q_2 i\gamma_5\tau_Q^2 Q_2\rangle \to \cos(2 \alpha)\, \langle
\bar Q_2 i\gamma_5\tau_Q^2 Q_2\rangle - \sin(2 \alpha)\,\langle \bar Q_2
i\gamma_5\tau_Q^1 Q_2\rangle\,;
\end{equation}
i.e., it is a ground state that is not invariant under the transformation.
Hence, for $(m=0,\mu\neq 0)$, only one Goldstone mode remains.  These
symmetry breaking patterns and the concomitant numbers of Goldstone modes in
QC$_2$D are also described in Ref.~\cite{stephanov3}.

\begin{figure}[t]
\centering{\
\epsfig{figure=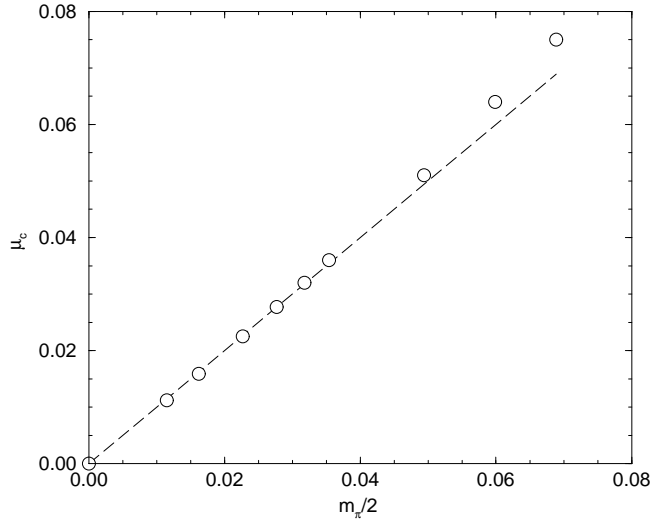,height=7.0cm}}
\caption{\label{mucm} Evolution of the critical chemical potential for
diquark condensation as the current-quark mass is increased.  The coordinate
measures the magnitude of the current-quark mass through the mass of the
theory's lightest excitation [a pseudo-Goldstone mode, as described after
Eq.~(\protect\ref{cndst})].  $\mu_c$ and $m_\pi$ are measured in units of the
model's mass scale, $\hat\eta$ in Eq.~(\protect\ref{delta}): for $m=0$, the
vector meson mass is $\mbox{\small$\frac{1}{\sqrt{2}}$}\hat\eta$.}
\end{figure}

For $m\neq 0$ and small values of $\mu$ the gap equation only admits a
solution with $\Delta \equiv 0$; i.e., diquark condensation is blocked
because the current-quark mass is a source of the quark condensate [see,
e.g., Eq.~(\ref{Bmeqn}) and the comments after Eq.~(\ref{cndst})].  However,
with increasing $\mu$, the theory undergoes a transition to a phase in which
the diquark condensate is nonzero.  We identify the transition as second
order because the diquark condensate falls continuously to zero as $\mu\to
\mu_c^+$, where $\mu_c$ is the critical chemical potential.  In
Fig.~\ref{mucm} we plot the critical chemical potential as a function of
$m_\pi/2$, where, to sidestep solving the Bethe-Salpeter equation, $m_\pi$
was obtained using a Gell-Mann--Oakes--Renner-like mass formula,
Eqs. (16)-(18) in Ref.~\cite{gregp}, which follows~\cite{fr96} from the
axial-vector Ward-Takahashi identity.  From the figure it is clear that this
simple model of QC$_2$D exhibits the relation
\begin{equation}
\label{mucmline}
\mu_c = \sfrac{1}{2}\,m_\pi\,,
\end{equation}
which is anticipated for QCD-like theories with pseudoreal
fermions~\cite{hands}.  We note that the deviation from
Eq.~(\protect\ref{mucmline}) at larger values of $m_\pi$ results from
neglecting O$(m^2)$-corrections in the mass formula.  This omission leads to
an underestimate of the pion mass~\cite{mr97}, which is responsible for the
upward deflection of the calculated results evident in
Fig.~\protect\ref{mucm}.

In exemplifying these features we have employed the rainbow-ladder
truncation.  However, improving on that will only yield quantitative changes
of $\lsim 20$\% in the results because the pseudoreality of QC$_2$D and the
equal dimension of the colour and bispinor spaces, which underly the theory's
Pauli-G\"ursey symmetry, ensure that the entire discussion remains
qualitatively unchanged. In particular, the results of Fig.~\ref{mucm} and
Eq.~(\ref{mucmline}), being tied to chiral symmetry, remain unchanged because
at least one systematic, chiral symmetry preserving truncation scheme
exists~\cite{truncscheme}.

\section{Three Colours}
The exploration of superfluidity in true QCD encounters two differences: the
dimension of the colour space is greater than that of the bispinor space and
the fundamental and conjugate representations of the gauge group are not
equivalent.  The latter is of obvious importance because it entails that the
quark-quark and quark-antiquark scattering matrices are qualitatively
different.

$n_c^\wedge=3$ in QCD and hence in canvassing superfluidity it is necessary
to choose a direction for the condensate in colour space;\footnote{It is this
selection of a direction in colour space that opens the possibility for
colour-flavour locked diquark condensation in a theory with three
effectively-massless quarks; i.e., current-quark masses $\ll
\mu$~\protect\cite{krishna,thomas}.} e.g., $\Delta^i\lambda^i_\wedge =
\Delta\, \lambda^2$ in Eq.~(\ref{sinv}), so that
\begin{equation}
\label{sinvQCD}
\BD(p,\mu) = \left(
\begin{array}{c|c}
D_\|(p,\mu) P_\| + D_\perp(p,\mu) P_\perp 
        & \Delta(p,\mu) \gamma_5 \lambda^2 \\ \hline
- \Delta(p,-\mu) \gamma_5 \lambda^2 \mbox{\rule{0mm}{1.0em}}
        & C D_\|(-p,\mu)C^\dagger P_\| + C D_\perp(p,\mu)C^\dagger P_\perp 
\end{array}\right)\,,
\end{equation}
where $P_\|=(\lambda^2)^2$, $P_\perp + P_\| = {\rm diag}({1,1,1})$, and
$D_\|$, $D_\perp$ are defined via obvious generalisations of
Eqs.~(\ref{sinv}), (\ref{DCA}).  In Eq.~(\ref{sinvQCD}) the evident,
demarcated block structure makes explicit the bispinor index: each block is a
$3\times 3$ colour matrix and the subscripts: $\|$, $\perp$, indicate whether
or not the subspace is accessible via $\lambda_2$.

The bispinors associated with this representation are given in
Eqs.~(\ref{QQCD}) and in this case the Lagrangian's quark-gluon interaction
term is
\begin{eqnarray}
\label{bispingamma}
\bar Q(x) i g\Gamma_\mu^a Q(x) A_\mu^a(x)\,,\;\; \Gamma_\mu^a & = &
\left(\begin{array}{c|c} \sfrac{1}{2}\gamma_\mu \lambda^a & 0
\\[0.2\parindent]\hline 0 \mbox{\rule{0mm}{1.0em}} &
-\sfrac{1}{2}\gamma_\mu (\lambda^a)^{\rm t}\end{array}\right)\,.
\end{eqnarray}
It is instructive to compare this with Eq.~(\ref{IQC2D}): with three colours
the interaction term is not proportional to the identity matrix in the
bispinor space, $\tau_Q^0$.  This makes plain the inequivalence of the
fundamental and conjugate fermion representations of $SU_c(3)$, which entails
that quark-antiquark scattering is different from quark-quark scattering.

It is straightforward to derive the gap equation at arbitrary order in the
truncation scheme of Ref.~\cite{truncscheme} and it is important to note that
because
\begin{eqnarray}
\lefteqn{
D_\|(p,\mu) P_\| + D_\perp(p,\mu) P_\perp = }\\
&& \nonumber \lambda^0 \left\{\sfrac{2}{3}
D_\|(p,\mu) + \sfrac{1}{3} D_\perp(p,\mu) \right\} +
\sfrac{1}{\sqrt{3}}\lambda^8 \left\{ D_\|(p,\mu) - D_\perp(p,\mu)\right\}
\end{eqnarray}
the interaction: $\Gamma_\mu^a {\cal S}(p,\mu)\Gamma_\nu^a$, necessarily
couples the $\|$- and $\perp$-components.  Reference~\cite{jacquesdq}
explored the possibility of diquark condensation in QCD using both the
rainbow and vertex-corrected gap equation, illustrated in Fig.~\ref{dressed},
with
\begin{equation}
g^2 D_{\mu\nu}(k) = \left(\delta_{\mu\nu}-\frac{k_\mu k_\nu}{k^2}\right)
                {\cal F}(k^2)\,,\; {\cal F}(k^2)= 4\pi^4\,
                \eta^2\,\delta^4(k)\,.
\label{delta3}
\end{equation}

For $(m,\mu)=0$ the rainbow-ladder truncation yields
\begin{equation}
\label{RLresults}
\begin{array}{ccc}
m_\omega^2 = m_\rho^2  =  \sfrac{1}{2}\,\eta^2\!,\; &
\langle \bar q q \rangle^0  = (0.11\,\eta)^3\!,\;&
{\cal B}_B(\mu=0)  =  (0.10\,\eta)^4\!,
\end{array}
\end{equation}
and momentum-dependent vector self energies, which lead to an interaction
between the $\|$- and $\perp$-components of $\BD$ that blocks diquark
condensation.  This is in spite of the fact that
$ \lambda^a \lambda^2 (-\lambda^a)^{\rm t} = \sfrac{1}{2}\lambda^a
\lambda^a, $
which entails that the rainbow-truncation quark-quark scattering kernel is
purely attractive and strong enough to produce diquark bound
states~\cite{justindq}. (Remember that in the colour singlet meson channel
the rainbow-ladder truncation gives the colour coupling $\lambda^a
\lambda^a$; i.e., an interaction with the same sign but twice as strong.)

For $\mu\neq 0$ and in the {\it absence} of diquark condensation this model
and truncation exhibits~\cite{bastiscm} coincident, first order chiral
symmetry restoring and deconfining transitions at
\begin{equation}
\mu_{c,\,{\rm rainbow}}^{B,\Delta=0} = 0.28 \,\eta = 0.3\,{\rm GeV}\,,
\end{equation}
with $\eta = 1.06\,$GeV fixed by fitting the $m\neq 0$ vector meson
mass~\cite{truncscheme}.

For $(m=0,\mu\neq 0)$, however, the rainbow-truncation gap equation admits a
solution with $\Delta(p,\mu)\not\equiv 0$ and $B(p,\mu)\equiv 0$.  The
pressure difference, $\delta p(\mu)$ in Eq.~(\ref{deltaP}), is again the way
to determine whether the stable ground state is the Nambu-Goldstone or
superfluid phase.  With increasing $\mu$, ${\cal B}_B(\mu)$ decreases, very
slowly at first, and ${\cal B}_{\Delta}(\mu)$ increases rapidly from zero.
That evolution continues until
\begin{equation}
\label{muc3}
\mu_{c,\,{\rm rainbow}}^{B=0,\Delta} = 0.25 \,\eta = 0.89\,\mu_{c,\,{\rm
rainbow}}^{B,\Delta=0}\,,
\end{equation}
where ${\cal B}_\Delta(\mu)$ becomes greater-than ${\cal B}_B(\mu)$.  This
signals a first order transition to the superfluid ground state and at the
boundary
\begin{equation}
\label{qqqbq}
\langle \bar Q i\gamma_5\tau_Q^2\lambda^2 Q\rangle_{\mu=\mu_{c,\,{\rm
rainbow}}^{B=0,\Delta}} 
= (0.65)^3\,\langle \bar Q Q\rangle_{\mu=0}\,.
\end{equation}
Since ${\cal B}_\Delta(\mu) > 0$ for all $\mu>0$ there is no intermediate
domain of $\mu$ in which all condensates vanish.

\begin{figure}[t]
\centering{\ \epsfig{figure=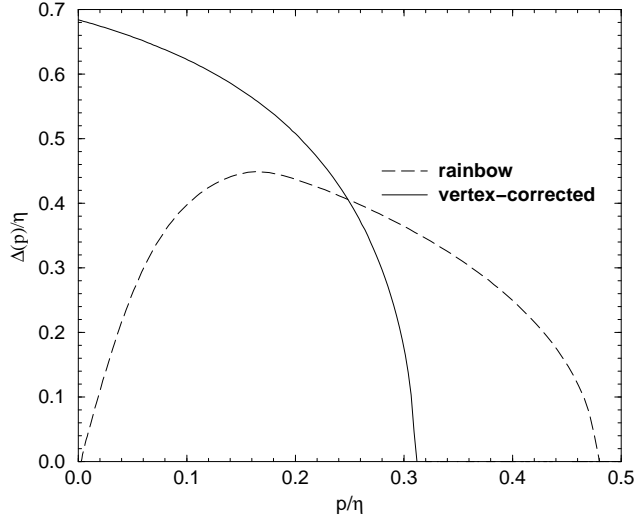,height=7.0cm}}
\caption{Dashed line: $\Delta(z,\mu_c^{B,\Delta})$ obtained in rainbow
truncation with the QCD model defined via Eq.~(\protect\ref{delta3}), plotted
for $\alpha = 0$ as a function of $p$, where $z= p\,(0,0,\sin\alpha,i\mu+
\cos\alpha )$.  As $\mu$ increases, the peak position shifts to larger values
of $p$ and the peak height increases.  Solid line: $\Delta(z,\mu=0)$ obtained
as the solution of Eq.~(\protect\ref{dseNO}), the vertex-corrected gap
equation, also with $\alpha=0$.  (Adapted from
Ref.~\protect\cite{jacquesdq}.)
\label{deltak}}
\end{figure}

The solution of the rainbow gap equation: $\Delta(p,\mu_c^{B,\Delta})$, which
is real and characterises the diquark gap, is plotted in
Fig.~\ref{deltak}. It vanishes at $p^2=0$ as a consequence of the
$\|$-$\perp$ coupling that blocked diquark condensation at $\mu=0$, and also
at large $p^2$, which is a manifestation of this simple model's version of
asymptotic freedom.

The chemical potential\footnote{We note that in a two flavour free-quark gas
at $\mu=0.3\,$GeV the baryon number density is $1.5\, \rho_0$, where
$\rho_0=0.16\,$fm$^{-3}$.  In the same system at $\mu=0.55\,$GeV the baryon
number density is $> 10\,\rho_0$.  For comparison, the central core density
expected in a $1.4\,M_\odot$ neutron star is
$3.6$-$4.1\,\rho_0$~\protect\cite{wiringa}.  Arguments valid at ``asymptotically
large'' quark chemical potential are therefore unlikely to be relevant to
experimentally or observationally accessible systems.} at which the switch to
the superfluid ground state occurs, Eq.~(\ref{muc3}), is consistent with
other estimates made using models comparable to the rainbow-truncation
class~\cite{krishna,thomas,blaschkebiel,diakonov2,bergesbiel}, as is the
large magnitude of the gap at this
point~\cite{krishna,thomas,blaschkebiel,diakonov2}.

A question that now arises is: How sensitive is this phenomenon to the nature
of the quark-quark interaction?  As we discussed in connection with
Eq.~(\ref{diquarkmasses}), the inhomogeneous dressed-ladder BSE exhibits
particle-like singularities in the $0^+$ diquark channels and such states do
not exist in the strong interaction spectrum.  Does diquark condensation
persist when a truncation of the gap equation is employed that does not
correspond to a BSE whose solutions exhibit diquark bound states?  The vertex
corrected gap equation,
\begin{eqnarray}
\label{dseNO}
\lefteqn{{\cal D}(p,\mu)  =  {\cal D}_0(p,\mu) 
}\\
&& \nonumber + \sfrac{3}{16}\eta^2\,\Gamma_\rho^a \,{\cal S}(p,\mu)\,
\Gamma_\rho^a - \sfrac{9}{256}\eta^4\,\Gamma_\rho^a\, {\cal S}(p,\mu)\,
\Gamma_\sigma^b\, {\cal S}(p,\mu)\, \Gamma_\rho^a\, {\cal S}(p,\mu)\,
\Gamma_\sigma^b\,,
\end{eqnarray}
which is depicted in Fig.~\ref{dressed}, is just such a truncation and it was
also studied in Ref.~\cite{jacquesdq}.

In this case there is a $\Delta \not \equiv 0 $ solution even for $\mu=0$,
which is illustrated in Fig.~\ref{deltak}, and using the interaction of
Eq.~(\ref{delta3})
\begin{equation}
\begin{array}{ccc}
m_\rho^2  =  (1.1)\, \,m_\rho^{2\;{\rm ladder}}, &
\; \langle \bar Q Q\rangle  =  (1.0)^3\,\langle \bar Q Q\rangle^{\rm
rainbow} ,&
\; {\cal B}_B  =  (1.1)^4\, {\cal B}_B^{\,{\rm rainbow}}\,,
\end{array}
\end{equation}
where the rainbow-ladder results are given in Eqs.~(\ref{RLresults}), and
\begin{equation}
\langle \bar Q i\gamma_5\tau_Q^2\lambda^2 Q\rangle = (0.48)^3\,\langle \bar Q
 Q\rangle\,,\;\; {\cal B}_\Delta = (0.42)^4\,{\cal B}_B \,.
\end{equation}
The last result shows, unsurprisingly, that the Nambu-Goldstone phase is
fa\-vour\-ed at $\mu=0$.  {\it Precluding} diquark condensation, the solution
of the vertex-corrected gap equation exhibits coincident, first order chiral
symmetry restoring and deconfinement transitions at
\begin{equation}
\mu_c^{B,\Delta=0} = 0.77\,\mu_{c,\,{\rm rainbow}}^{B,\Delta=0}\,.
\end{equation}

Admitting diquark condensation, however, the $\mu$-dependence of the bag
constants again shows there is a first order transition to the superfluid
phase, here at
\begin{equation}
\mu_{c}^{B=0,\Delta} = 0.63\, \mu_{c}^{B,\Delta=0}\,,\;{\rm with}\;
\langle \bar Q i\gamma_5\tau_Q^2\lambda^2 Q \rangle_{\mu =
0.63\,\mu_{c}^{B,\Delta=0}} = (0.51)^3\,\langle \bar Q Q\rangle_{\mu=0}\,.
\end{equation}
(NB.\ This discussion is still for $m=0$.  We saw at the end of
Sec.~\ref{sec:Nc2} what effects to anticipate at $m\neq 0$.)  Thus the
material step of employing a truncation that eliminates diquark bound states
leads only to small quantitative changes in the quantities characterising the
still extant superfluid phase; e.g., reductions in the magnitude of both the
critical chemical potential for the transition to superfluid quark matter and
the gap.  Hence scalar diquark condensation appears to be a robust
phenomenon.  One caveat to bear in mind, however, is that the gap equation
studies conducted hitherto do not obviate the question of whether the diquark
condensed phase is stable with-respect-to dinucleon
condensation~\cite{birse}, which requires further attention.

Heating causes the diquark condensate to evaporate.  Existing studies suggest
that it will disappear for $T \gsim
60$--$100\,$MeV~\cite{krishna,thomas,blaschkebiel,bergesbiel}.  However, such
temperatures are high relative to that anticipated inside dense astrophysical
objects, which may indeed therefore provide an environment for detecting
quark matter superfluidity.

\section{Summary}
The idea that diquark correlations play an important role in strong
interaction physics is an old one.  However, modern computational resources
and theoretical techniques make possible a more thorough and quantitative
exploration of the merits of this idea and its realisation in QCD.  These
advances are in part responsible for the contemporary resurgence of interest
in all aspects of diquark-related phenomena.

Herein we have attempted to provide a qualitative understanding of the nature
of diquark condensation using exemplary, algebraic models, and focusing on
two flavour theories for simplicity.

The gap equation is a primary tool in all studies of pairing.  Using the
special case of $2$-colour QCD, QC$_2$D, we illustrated via an analysis of
the gap equation how a nonzero chemical potential promotes Cooper pairing and
how that pairing can overwhelm a source for quark-antiquark condensation,
such as a fermion current-mass.  As we saw, the pseudoreality of $SU(N_c=2)$
entails that QC$_2$D has a number of special symmetry properties, which
dramatically affect the spectrum.

Turning to QCD itself, we saw that one can expect a nonzero quark condensate
at zero chemical potential: $\sigma:=-\langle \bar q q \rangle^{1/3} \neq 0$,
to give way to a diquark condensate when the chemical potential exceeds
$\approx 2 \sigma$, and at this point the diquark gap is $\approx \sigma/2$.
The diquark condensate melts when the temperature exceeds $\sim
60\,$--$\,100\,$MeV; i.e., one-third to one-half of the chiral symmetry
restoring temperature in two-flavour QCD.  These features are
model-independent in the sense that the many, disparate models applied
recently to the problem yield results in semi-quantitative agreement.

\bigskip

\hspace*{-\parindent}\parbox{34.5em}{{\large\bf
Acknowledgments}\\[\baselineskip]
This work was supported by the US Department of Energy, Nuclear Physics
Division, under contract no. W-31-109-ENG-38 and the National Science
Foundation under grant no.\ INT-9603385.  S.M.S. is grateful for financial
support from the A.v.~Humboldt foundation.}


\begin{thebibliography}{8.}
\addcontentsline{toc}{section}{References}
%
\bibitem{firstdq} M.~Ida and R.~Kobayashi, Prog.\ Theor.\ Phys.\ {\bf 36}
(1966) 846; D.B.~Lichtenberg and L.J.~Tassie, Phys.\ Rev.\ {\bf 155} (1967)
1601. 
%
\bibitem{lichtenberg} D.B.~Lichtenberg, ``Why Is It Necessary To Consider
Diquarks,'' in Proc.\ of the Workshop on Diquarks, edited by Mauro Anselmino
and Enrico Predazzi (World Scientific, Singapore, 1989) pp. 1-12.
%
\bibitem{L2} D.B.~Lichtenberg, L.J.~Tassie and P.J.~Keleman, Phys.\ Rev.\
{\bf 167} (1968) 1535.
%
\bibitem{L3} J.~Carroll, D.B.~Lichtenberg and J.~Franklin, Phys.\ Rev.\ {\bf
174} (1968) 1681.
%
\bibitem{regdq} R.T.~Cahill, J.~Praschifka and C.J.~Burden,
Austral. J. Phys. {\bf 42} (1989) 161;
R.T.~Cahill, {\it ibid} 171.
%
\bibitem{njl} S.P.~Klevansky, Rev.\ Mod.\ Phys.\ {\bf 64} (1992) 649;
G.~Ripka and M.~Jaminon, Annals Phys.\ {\bf 218} (1992) 51.
%
\bibitem{gcm98} P.C.~Tandy, Prog.\ Part.\ Nucl.\ Phys.\ {\bf 39} (1997) 117;
R.T.~Cahill and S.M.~Gunner, Fizika {\bf B 7} (1998) 171.
%
\bibitem{cdragw} C.D.~Roberts and A.G.~Williams, Prog.\ Part.\ Nucl.\ Phys.\
{\bf 33} (1994) 477.
%
\bibitem{revbasti} C.D.~Roberts and S.M.~Schmidt, Prog.\ Part.\ Nucl.\ Phys.\
{\bf 45} (2000) S1.
%
\bibitem{revreinhard} R.~Alkofer and L.~v.~Smekal, ``The infrared behavior of
QCD Green's functions: Confinement, dynamical symmetry breaking, and hadrons
as relativistic bound states,'' hep-ph/0007355.
%
\bibitem{pmqciv} P.~Maris, ``Continuum QCD and light mesons,''
nucl-th/0009064.
%
\bibitem{cdrqciv} M.B.~Hecht, C.D.~Roberts and S.M.~Schmidt, ``Contemporary
applications of Dyson-Schwinger equations,'' nucl-th/0010024.
%
\bibitem{justindq} R.T.~Cahill, C.D.~Roberts and J.~Praschifka, Phys.\ Rev.\
{\bf D 36} (1987) 2804.
%
\bibitem{truncscheme} A.~Bender, C.D.~Roberts and L.~v.~Smekal, Phys.\ Lett.\
{\bf B 380} (1996) 7.
%
\bibitem{reinharddq} G.~Hellstern, R.~Alkofer and H.~Reinhardt, Nucl.\ Phys.\
{\bf A 625} (1997) 697.
%
\bibitem{conradsep} C.J.~Burden, L.~Qian, C.D.~Roberts, P.C.~Tandy and
M.J.~Thomson, Phys.\ Rev.\ {\bf C 55} (1997) 2649.
%
\bibitem{latticediquark} M.~Hess, F.~Karsch, E.~Laermann and I.~Wetzorke,
Phys.\ Rev.\ {\bf D 58} (1998) 111502.
%
\bibitem{regfe} R.T.~Cahill, C.D.~Roberts and J.~Praschifka, Austral.\ J.\
Phys.\ {\bf 42} (1989) 129.
%
\bibitem{conradfe} C.J.~Burden, R.T.~Cahill and J.~Praschifka,
Austral.\ J.\ Phys.\  {\bf 42} (1989) 147.
%
\bibitem{reinhard} M.~Oettel, G.~Hellstern, R.~Alkofer and H.~Reinhardt,
Phys.\ Rev.\  {\bf C 58} (1998) 2459.
%
\bibitem{reinhardDelta} M.~Oettel, R.~Alkofer and L.~v~Smekal, Eur.\ Phys.\
J.\ {\bf A 8} (2000) 553. \label{reinhardDeltaR}
%
\bibitem{wallyexp} G.G.~Petratos, I.R.~Afnan, F.~Bissey, J.~Gomez,
A.T.~Katramatou, W.~Melnitchouk and A.W.~Thomas, ``Measurement of the
$F_2^{n}/F_2^{p}$ and $d/u$ Ratios in Deep Inelastic Electron Scattering off
$^3$H and $^3$He,''  nucl-ex/0010011.
%
\bibitem{kahana} D.~Kahana and U.~Vogl, Phys. Lett. {\bf B 244} (1990) 10.
%
\bibitem{bl84} D.~Bailin and A.~Love, Phys. Rept. {\bf 107} (1984) 325; and
references therein.
%
\bibitem{krishna} M.~Alford, J.~A.~Bowers and K.~Rajagopal, ``Color
superconductivity in compact stars,'' hep-ph/0009357, this volume.
%
\bibitem{thomas} T.~Sch\"afer and E.~Shuryak, ``Phases of QCD at high baryon
density,'' nucl-th/0010049, this volume.
%
\bibitem{fridolin} D.~Blaschke, H.~Grigorian, N.~Glendenning, G.~Poghosyan
and F. Weber, ``Deconfinement signals from Rotating Compact Stars,'' this
volume.
%
\bibitem{bastiscm} D.~Blaschke, C.D.~Roberts and S.M.~Schmidt, Phys.\ Lett.\
{\bf B 425} (1998) 232.
%
\bibitem{gregp} A.~Bender, G.I.~Poulis, C.D.~Roberts, S.M.~Schmidt and
A.W.~Thomas, Phys.\ Lett.\ {\bf B 431} (1998) 263.
%
\bibitem{arne2} A.~H\"oll, P.~Maris and C.D.~Roberts, Phys.\ Rev.\ {\bf C 59}
(1999) 1751.
%
\bibitem{jacquesdq} J.C.R.~Bloch, C.D.~Roberts and S.M.~Schmidt, Phys.\ Rev.\
{\bf C 60} (1999) 065208.
%
\bibitem{mattuck} R.D.~Mattuck: \emph{A Guide to Feynman Diagrams in the
Many-Body Problem} (McGraw-Hill, New York 1976).
%
\bibitem{cr85} R.T.~Cahill and C.D.~Roberts, Phys.\ Rev.\ {\bf D 32} (1985)
2419.
%
\bibitem{mn83} H.J.~Munczek and A.M.~Nemirovsky, Phys.\ Rev.\ {\bf D 28}
(1983) 181.
%
\bibitem{latticequark} J.I.~Skullerud and A.G.~Williams, ``Quark propagator
in Landau gauge,'' hep-lat/0007028.
%
\bibitem{quarkstar1} D.~Blaschke, H.~Grigorian, G.~Poghosyan, C.D.~Roberts
and S.M.~Schmidt, Phys.\ Lett.\ {\bf B 450} (1999) 207.
%
\bibitem{stephanov3} J.B.~Kogut, M.A.~Stephanov and D.~Toublan, Phys.\ Lett.\
{\bf B 464} (1999) 183.
%
\bibitem{fr96} M.R.~Frank and C.D.~Roberts, Phys.\ Rev.\ {\bf C 53} (1996)
390.
%
\bibitem{hands} S.~Hands and S.~Morrison, ``Diquark condensation in dense
matter: A Lattice perspective,'' in Proc. of the International Workshop on
Understanding Deconfinement in QCD, edited by D.~Blaschke, F.~Karsch and
C.D.~Roberts (World Scientific, Singapore, 2000) pp. 31-42;
%
J.B.~Kogut, M.A.~Stephanov, D.~Toublan, J.J.~Verbaarschot and
A.~Zhitnitsky, Nucl.\ Phys.\ {\bf B 582} (2000) 477;
%
R.~Aloisio, V.~Azcoiti, G.~Di Carlo, A.~Galante and
A.~F.~Grillo, ``Fermion condensates in two colours finite density QCD at
strong coupling,'' hep-lat/0009034;
R.~Aloisio,~V. Azcoiti, G.~Di Carlo, A.~Galante and A.F.~Grillo,
``Probability Distribution Function of the Diquark Condensate in Two Colours
QCD,'' hep-lat/0011079.
%
\bibitem{mr97} P.~Maris and C.D.~Roberts, Phys.\ Rev.\ {\bf C 56} (1997)
3369.
%
\bibitem{wiringa} R. B. Wiringa, V. Fiks and A. Fabrocini, Phys. Rev. {\bf C
38} (1988) 1010.
%
\bibitem{blaschkebiel} D.~Blaschke and C.D.~Roberts, Nucl.\ Phys.\ {\bf A
642} (1998) 197c.
%
\bibitem{diakonov2} G.W.~Carter and D.~Diakonov, ``Chiral symmetry breaking
and color superconductivity in the Instanton picture,'' in Proc. of the
International Workshop on Understanding Deconfinement in QCD, edited by
D.~Blaschke, F.~Karsch and C.D.~Roberts (World Scientific, Singapore, 2000)
pp. 239-250.
%
\bibitem{bergesbiel} J.~Berges, Nucl.\ Phys.\ {\bf A 642} (1998) 51c.
%
\bibitem{birse} S.~Pepin, M.~C.~Birse, J.A.~McGovern and N.R.~Walet, Phys.\
Rev.\ {\bf C 61} (2000) 055209.
%
\end{thebibliography}
\end{document}